\documentclass[twocolumn,showpacs,aps,prd,superscriptaddress,floatfix]{revtex4}
\usepackage{graphicx}
\usepackage{lineno}
\usepackage{dcolumn}
\usepackage{bm}

\newcommand{\BaBarYear}       {07}
\newcommand{\BaBarNumber}     {007}
\newcommand{\SLACPubNumber} {0000}

\newcommand{\BaBarType}      {PUB}  

\input babarsym.tex

\def\Vub {\ensuremath{V_{ub}}}

\def\btn {\ensuremath{B^{+} \to \tau^{+} \nu}}
\def\btotaunu {\ensuremath{B^{+} \to \tau^{+} \nu}}
\def\btodlnux {\ensuremath{\Bub \to \Dz \ell^{-} \bar{\nu}_{\ell} X}}

\def\eextra {\ensuremath{E_{\mathrm{extra}}}}
\def\mmiss  {\ensuremath{M_{\mathrm{miss}}}}

\def\nulb  {\ensuremath{\bar{\nu}_{\ell}}}
\def\nueb  {\ensuremath{\bar{\nu}_{e}}}

\def\nutb  {\ensuremath{\bar{\nu}_{\tau}}}

\def\tautoenunu {\ensuremath {\tau^+ \to e^+ \nu \nub}}
\def\tautoe {\ensuremath {\tau^+ \to e^+ \nu \nub}}
\def\enunu {\ensuremath {e^+ \nu \nub}}
\def\tautomununu {\ensuremath {\tau^+ \to \mu^+ \nu \nub}}

\def\mununu {\ensuremath {\mu^+ \nu \nub}}
\def\tautopinu {\ensuremath {\tau^+ \to \pi^+ \nub}}
\def\tautopi {\ensuremath {\tau^+ \to \pi^+ \nub}}
\def\pinu {\ensuremath {\pi^+ \nub}} 
\def\tautopipiznu {\ensuremath {\tau^+ \to \pi^+ \pi^{0} \nub}}

\def\tautorho {\ensuremath {\tau^+ \to \pi^+ \pi^{0} \nub}}
\def\pipiznu {\ensuremath {\pi^+ \pi^{0} \nub}}

\def\tautothreepi {\ensuremath {\tau^+ \to \pi^+ \pi^{-} \pi^{+} \nub}}

\def\bcount    {\ensuremath {383 \times 10^{6}} }
\def\onlumi    {\ensuremath { 346  \invfb\ }}
\def\offlumi   {\ensuremath { 36.3 \invfb\  }}

\long\def\inst#1{\par\nobreak\kern 4pt\nobreak
    {\it #1}\par\vskip 10pt plus 3pt minus 3pt}

\begin{document}

{\pagestyle{empty}

\begin{flushleft}
\babar-\BaBarType-\BaBarYear/\BaBarNumber \\
SLAC-PUB-\SLACPubNumber
\end{flushleft}

\title{
  {\bf \boldmath
   A Search for $\btotaunu$}
}

%
\author{B.~Aubert}
\author{M.~Bona}
\author{D.~Boutigny}
\author{Y.~Karyotakis}
\author{J.~P.~Lees}
\author{V.~Poireau}
\author{X.~Prudent}
\author{V.~Tisserand}
\author{A.~Zghiche}
\affiliation{Laboratoire de Physique des Particules, IN2P3/CNRS et Universit\'e de Savoie, F-74941 Annecy-Le-Vieux, France }
\author{J.~Garra~Tico}
\author{E.~Grauges}
\affiliation{Universitat de Barcelona, Facultat de Fisica, Departament ECM, E-08028 Barcelona, Spain }
% ...
\collaboration{The \babar\ Collaboration}
\noaffiliation

\date{\today}

\begin{abstract}
\noindent We present a search for the decay \btn\ using \bcount $\B\Bbar$ pairs   
collected at the $\Y4S$ resonance with the \babar\ detector 
at the SLAC PEP-II $B$-Factory. A sample of events with one reconstructed 
semileptonic $B$ decay (\btodlnux) is selected, and
in the recoil a search for $\btn$ is performed. The $\tau$ is
identified in the following channels: $\tautoenunu$, $\tautomununu$,
$\tautopinu$ and $\tautopipiznu$. 
We measure a branching fraction of 
$\mathcal{B}(\btn)=(0.9\pm{0.6}(\mbox{stat.}) \pm 0.1 (\mbox{syst.})) \times 10^{-4}$. 
In the absence of a significant signal, we calculate an upper limit at the 90\% confidence 
level of $\mathcal{B}(\btn) < 1.7 \times 10^{-4}$.
We calculate the product of the $B$ meson decay constant $f_{B}$ and $|\Vub|$ to be
$f_{B}\cdot|\Vub| = (7.2^{+2.0}_{-2.8}(\mbox{stat.})\pm{0.2}(\mbox{syst.}))\times10^{-4}$~GeV.
\end{abstract}


\vfill

\maketitle
}

\section{\label{sec:level1}Introduction}

In the Standard Model (SM), the purely leptonic decay $\btn$~\citep*{cc} 
proceeds via quark annihilation into a $W^{+}$ boson.
The branching fraction is given by:
\begin{equation}
\label{eqn:br}
\mathcal{B}(B^{+} \rightarrow {\taup} \nu)= 
\frac{G_{F}^{2} m^{}_{B}  m_{\tau}^{2}}{8\pi}
\left[1 - \frac{m_{\tau}^{2}}{m_{B}^{2}}\right]^{2} 
\tau_{\Bu} f_{B}^{2} |\Vub|^{2},
\end{equation}
where  
$\Vub$ is an element of the Cabibbo-Kobayashi-Maskawa quark-mixing 
matrix~\citep*{c,km}, 
$f_{B}$ is the $B$ meson decay constant, 
$G_F$ is the Fermi constant,
$\tau_{\Bu}$ is the $\Bu$ lifetime, and
$m^{}_{B}$ and $m_{\tau}$ are the $\Bu$ and $\tau$ masses.
Physics beyond the SM, such as two-Higgs doublet models,
could enhance or suppress $\mathcal{B}(\btn)$ through the
introduction of a charged Higgs boson~\cite{higgs,Isidori2006pk,Akeroyd2007eh}.
Using theoretical calculations
of $f_B$ from lattice QCD and experimental measurements of $|\Vub|$
from semileptonic $B$ decays, this purely leptonic $B$ decay can be used
to constrain the parameters of theories beyond the SM.
Or, assuming that SM processes dominate and using the value 
of $|\Vub|$ determined from semileptonic $B$ decays,
purely leptonic decays provide an experimental method of
measuring $f_B$ with reduced theoretical error.

The branching fractions for $B^+\to\mu^+\nu$ and $B^+\to e^+\nu$ are suppressed 
by factors of $\sim 5\times 10^{-3}$ and $\sim 10^{-7}$ with respect to $\btn$. 
However, a search for $\btn$ is experimentally challenging 
due to the large missing momentum from multiple neutrinos, which makes 
the signature less distinctive than in the other leptonic modes.
The SM estimate of the branching fraction for $\btn$,
using $|\Vub| = (4.31 \pm 0.30)\times 10^{-3}$~\cite{pdg2004} and 
$f_{B} = 0.216 \pm 0.022$ GeV~\citep*{fb} in Eq.~\ref{eqn:br}
is $(1.6 \pm 0.4)\times 10^{-4}$.
In a previously published analysis using a sample of 
$223 \times 10^6$ $\FourS$ decays,
the \babar\ collaboration set an upper limit of
$\mathcal{B}(\btn) < 2.6 \times 10^{-4} \, 
\textrm{ at the 90\% confidence level (CL)}$~\citep{babar-prd-btn}.
The Belle Collaboration has reported evidence from a search for this channel where the branching fraction was measured to be
$\mathcal{B}(\btn) = (1.79^{+0.56}_{-0.49}(\mbox{stat.}) ^{+0.46}_{-0.51}(\mbox{syst.})) \times 10^{-4}$
~\citep{belle}.

\section{The \boldmath{\babar} Detector and Dataset}
\label{sec:babar}
The data used in this analysis were collected with the \babar\ detector~\citep{babar}
at the \pep2\ storage ring. 
The sample corresponds to an integrated
luminosity of \onlumi at the \FourS\ resonance (on-resonance) 
and \offlumi taken at $40\mev$ below the $\BB$ production threshold 
(off-resonance) which is used to study background from
$e^+e^-\to f\bar{f}$ ($f = u, d, s, c, \tau$) continuum events. 
The on-resonance sample contains $(383 \pm 4)\times 10^{6}$  $\FourS$ decays. 
The detector components used in this analysis are the tracking system
composed of a five-layer silicon vertex detector and a 40-layer drift chamber (DCH),
the Cherenkov detector for charged $\pi$--$K$ discrimination, a CsI calorimeter
(EMC) for photon and electron identification, and an 
18--layer flux return (IFR) located outside of the 1.5~T solenoidal coil 
and instrumented with resistive plate chambers for muon
and neutral hadron identification. 
For the most recent 133\invfb of data, a portion of the 
resistive plate chambers
has been replaced with limited streamer tubes.
We analyze the data from
several data-taking periods separately to account for
varying accelerator and detector conditions.

A GEANT4-based \cite{geant4} Monte Carlo (MC) 
simulation is used to model signal efficiencies and physics backgrounds. 
The $\tau$ lepton decay is modeled using EvtGen \cite{evtgen}.
Beam-related background and detector 
noise from data are overlaid on the simulated events.
Simulation samples equivalent to approximately three times the accumulated data  are
used to model \BB\ events, and samples equivalent to approximately
1.5 times the accumulated data are used to model continuum events.
We determine selection efficiencies for signal events using a 
MC simulation where one $B^+$ meson decays to $\tau^+\nu$, 
while the other is allowed to decay into any final state.

\section{ANALYSIS METHOD} 
\label{sec:Analysis}

Due to the presence of multiple neutrinos, the \btn \xspace decay mode
lacks the kinematic constraints which are usually exploited in $B$ decay 
searches in order to reject both continuum and \BB\ backgrounds.
The strategy adopted for this analysis is to reconstruct exclusively 
the decay of one of the $B$ mesons in the event, referred to as the 
``tag'' $B$. The remaining particle(s) in the event (the ``recoil'') are assumed to
come from the other $B$ and are compared with the signature
expected for \btn. In order to avoid experimenter bias, the 
signal region in data is  blinded until the final yield
extraction is performed.

The tag $B$ is reconstructed in the set of semileptonic $B$ decay modes 
\btodlnux, where $\ell$ denotes either electron or muon, and $X$ can be either nothing or a 
transition particle ($\piz$ or photon) from a higher mass charm state decay which we do not 
attempt to explicitly include in the tag $B$. 
However, we explicitly veto events where the best tag candidate
is consistent with neutral $B$ decay, where the $X$ system is a single charged pion
that can be combined with the \Dz\ to form a \Dstarp\ candidate.

The \btn \xspace signal is searched for in
both leptonic and hadronic $\tau$ decay modes constituting approximately 
71\% of the total $\tau$ decay width:
$\tautoenunu$, $\tautomununu$, $\tautopinu$, and $\tautopipiznu$.
We do not consider the $\tautothreepi$ mode since we found
it to be dominated by background events.

\subsection{Tag \boldmath{$B$} Reconstruction}
\label{sec:TagReco}

\begin{figure}
\begin{center}
\includegraphics[width=0.5\linewidth,keepaspectratio]{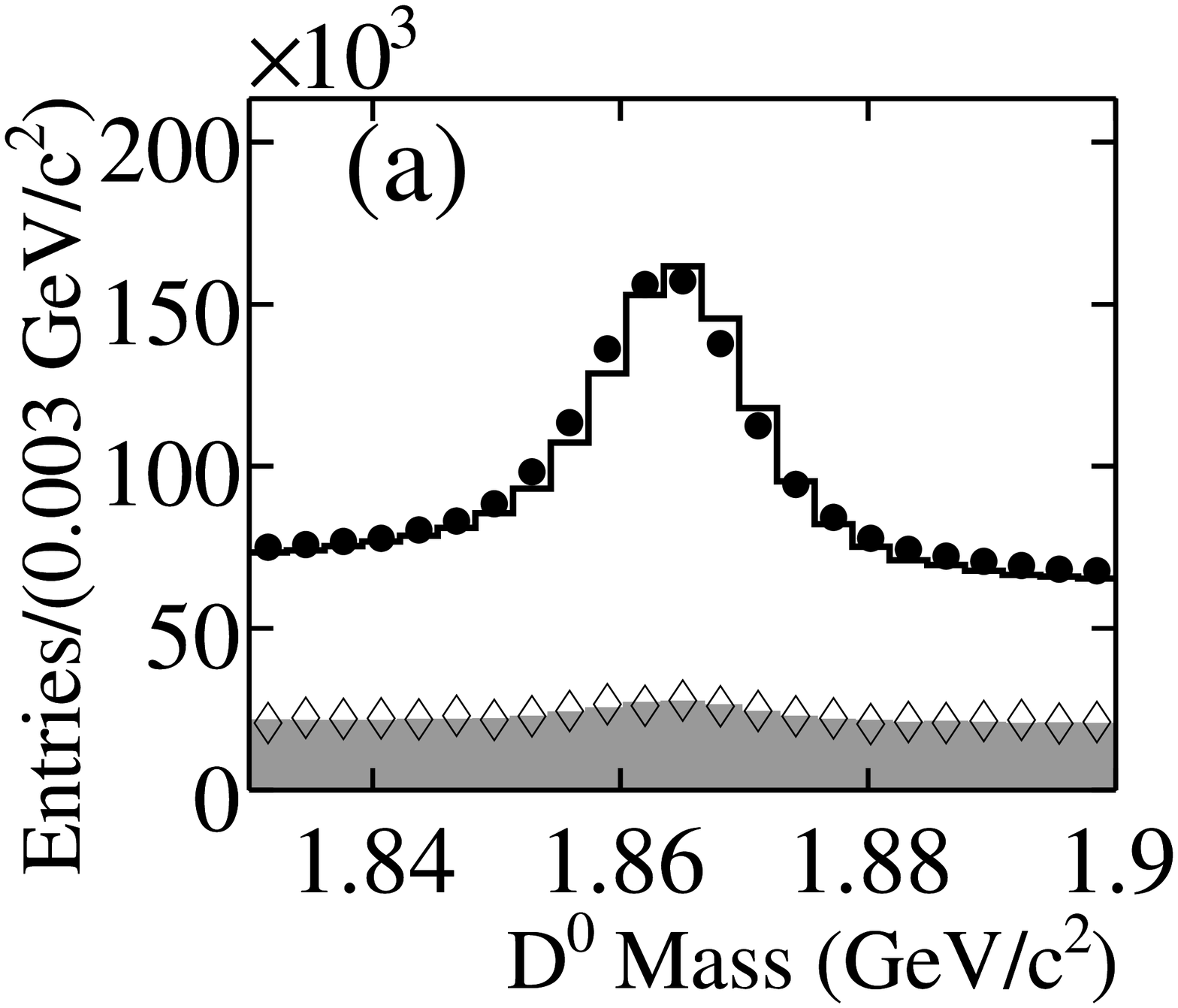}%
\includegraphics[width=0.5\linewidth,keepaspectratio]{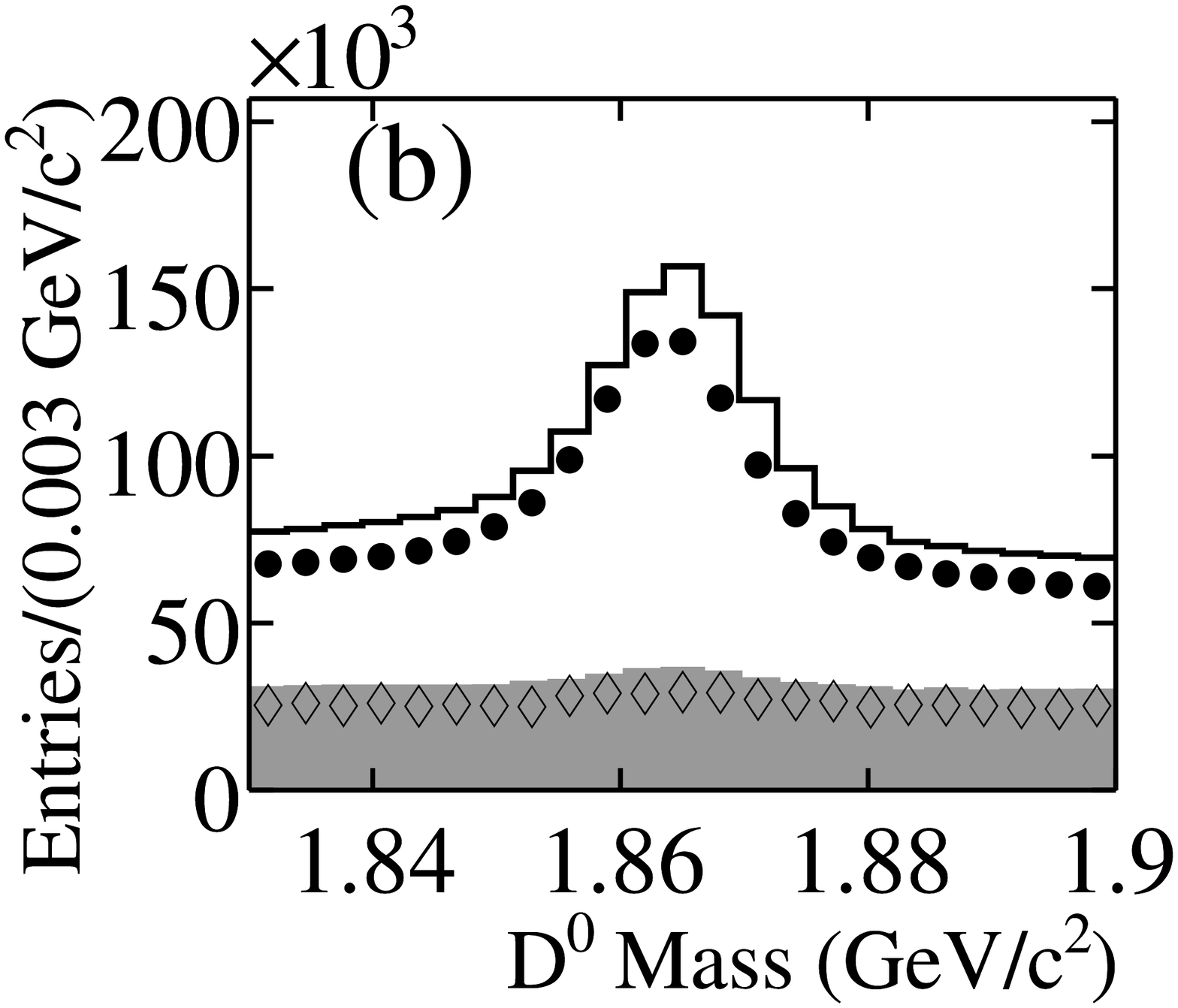}

\includegraphics[width=0.5\linewidth,keepaspectratio]{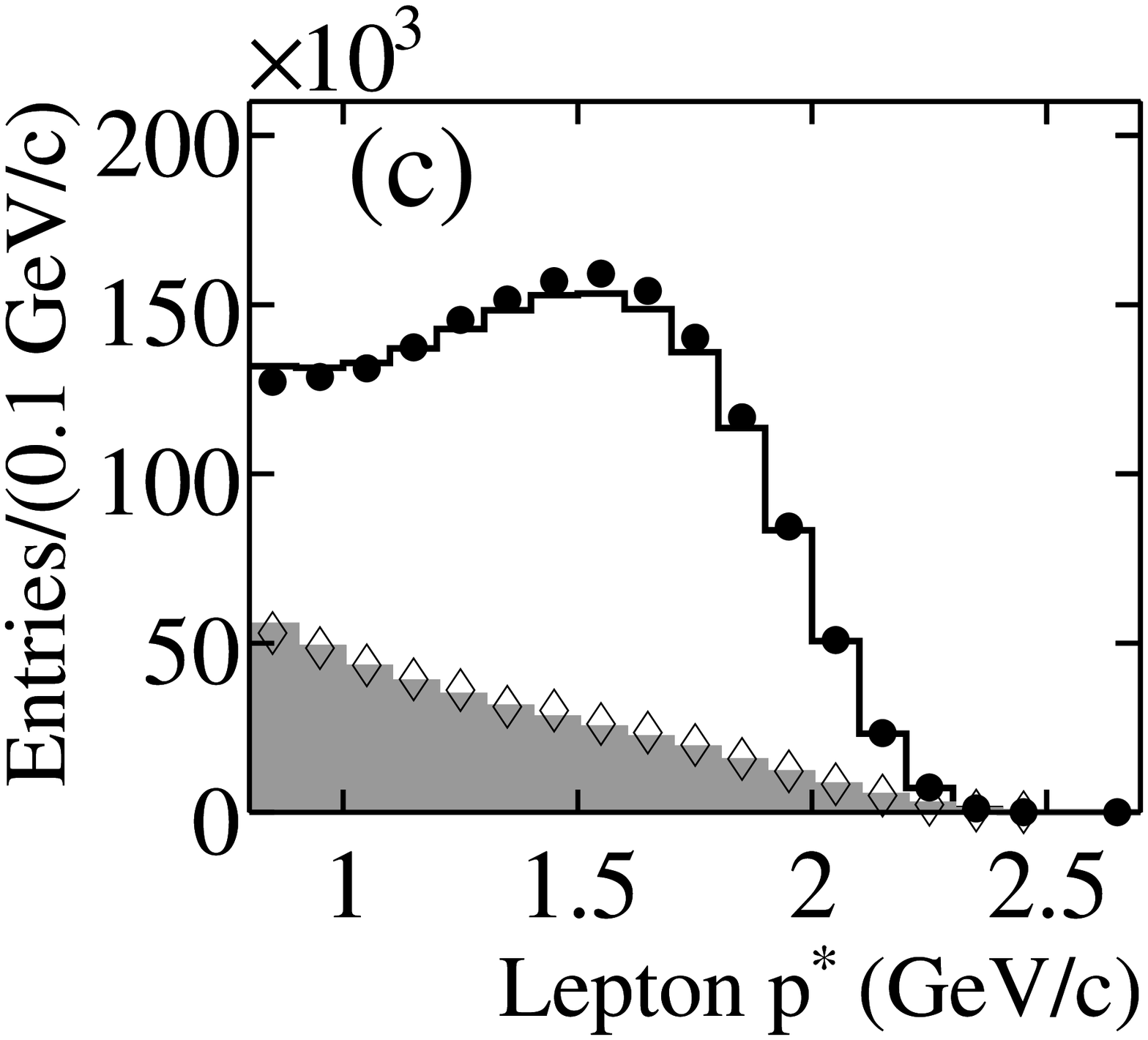}%
\includegraphics[width=0.5\linewidth,keepaspectratio]{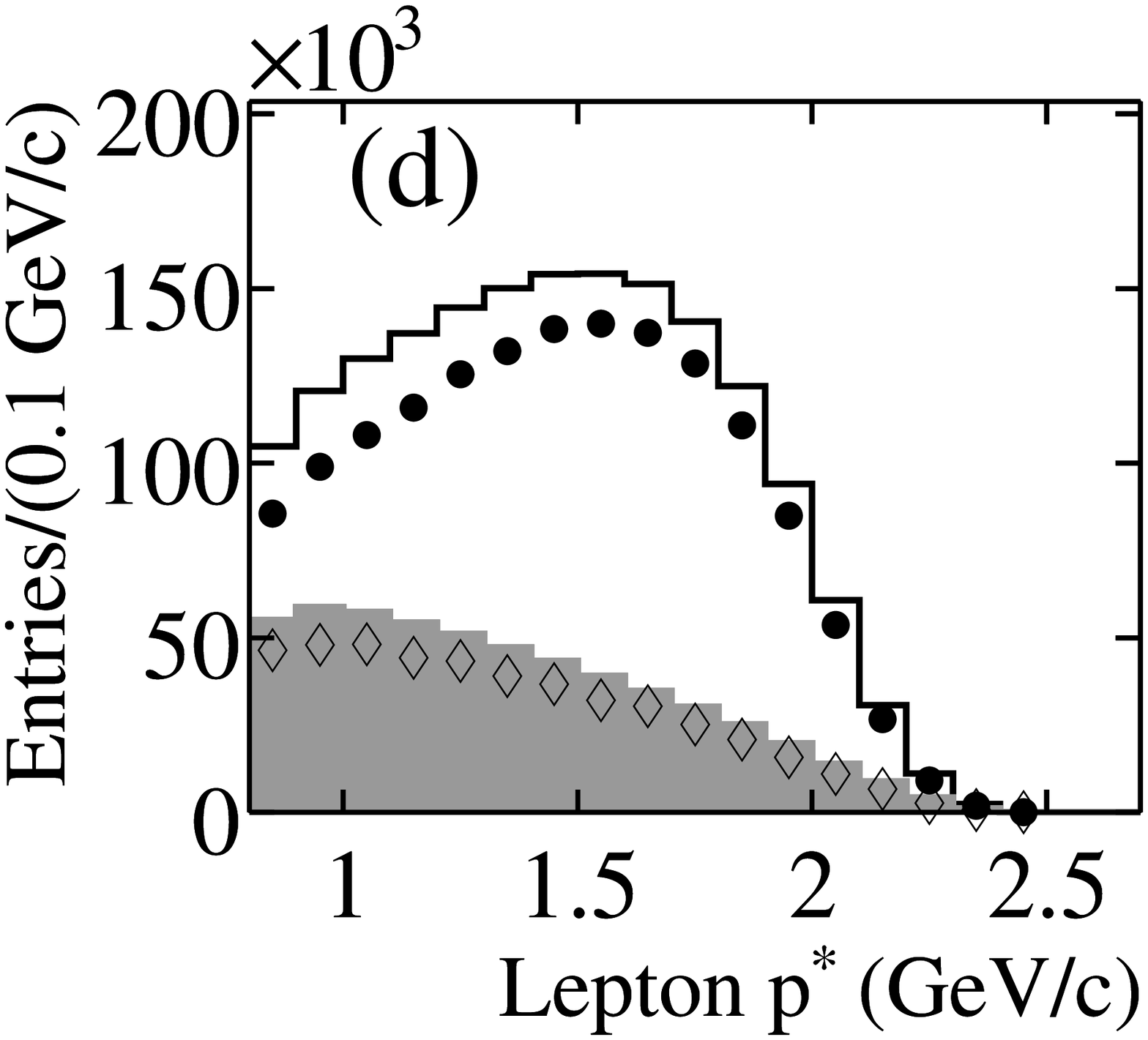}
\end{center}
\vspace{-0.30cm}
\caption{\Dz\ mass for tag $B$ candidates containing an (a) electron or (b) muon and the CM momentum of the tag $B$ lepton for tag $B$ candidates containing an (c) electron or (d) muon.
On-resonance data (filled circles) are overlaid on the \BB\ MC (solid histogram) and non-resonant background MC (gray histogram), which have been normalized to the integrated data luminosity. 
Off-resonance data (open diamonds) are overlaid for comparison, and normalized to the on-resonance integrated luminosity. 
}
\label{fig:d0masstag_pstartaglep}
\end{figure}

$\Dz\ell$ candidates are reconstructed by combining a
$\Dz$ with an identified electron 
or muon with momentum above $0.8\gevc$ in the $e^+e^-$ center-of-mass (CM) frame (Fig.~\ref{fig:d0masstag_pstartaglep}). 
The flight direction of the $\Dz$ is required to intersect with
the lepton track.
Assuming that the massless neutrino is the only missing particle, we 
calculate the cosine of the angle between the $\Dz\ell$ candidate
and the $B$ meson,
\begin{equation}
\cos\theta_{B-\Dz\ell} = \frac{2 E_{B} E_{\Dz\ell} - m_{B}^{2} - 
m_{\Dz\ell}^{2}}{2|\vec{p}_{B}||\vec{p}_{\Dz\ell}|},
\label{eqn:cosby}
\end{equation}
where ($E_{\Dz\ell}$, $\vec{p}_{\Dz\ell}$) and
($E_{B}$, $\vec{p}_{B}$) are the 
four-momenta of the $\Dz\ell$ and $B$ in the CM frame, and $m_{\Dz\ell}$ and $m_{B}$ 
are the masses of the $\Dz\ell$ candidate and $\Bu$ meson (the nominal mass~\cite{pdg2004} is used), respectively. 
$E_{B}$ and the magnitude of $\vec{p}_{B}$ are calculated 
from the beam energy: $E_{B} = E_{\rm{CM}}/2$, where $E_{\rm{CM}}$ is the CM energy of the beams, and 
$ | \vec{p}_{B} | = \sqrt{E_{B}^{2} - m_{B}^{2} }$.
Correctly reconstructed candidates
populate $\cos\theta_{B-\Dz\ell}$ in the range of [$-1,1$], whereas combinatorial backgrounds
can take unphysical values outside this range. 
We retain events in the interval 
$-2.0 < \cos\theta_{B-\Dz\ell} < 1.1$, where the upper bound takes 
into account the detector resolution and the less restrictive lower bound
accepts those events where the $X$ is a soft transition particle from a higher mass
charm state.

We reconstruct the $\Dz$ candidates in four decay modes:
$K^{-}\pi^{+}$, $K^{-}\pi^{+}\pi^{-}\pi^{+}$, $K^{-}\pi^{+}\pi^{0}$, and
$\KS\pi^{+}\pi^{-}$, only considering $\KS$ candidates decaying to charged pions. 
The charged tracks are required to meet particle identification criteria consistent
with the particle hypothesis and are required to converge at a common vertex.
The $\piz$ candidates are required to have invariant masses between
0.115 and 0.150$\gevcc$ and the photon daughter candidates of the $\piz$ must 
have a minimum laboratory energy of 30$\mev$ and have shower shapes consistent with 
electromagnetic showers. 
The mass of the reconstructed $\Dz$ candidates (Fig.~\ref{fig:d0masstag_pstartaglep}) in the
$K^{-}\pi^{+}$, $K^{-}\pi^{+}\pi^{-}\pi^{+}$, and $\KS\pi^{+}\pi^{-}$
modes is required to be within 20$\mevcc$ of the nominal mass 
\cite{pdg2004}, while in the $K^{-}\pi^{+}\pi^{0}$ decay mode 
the mass is required to be within 35$\mevcc$ of the nominal mass. 
These constraints are determined by fitting a single Gaussian function and
a first-order polynomial to the mass distribution in signal MC and correspond to the
$3\sigma$ positions on the Gaussian.
Furthermore, the sum of the charges of all the particles 
in the event must be equal to zero.
If more than one suitable $\Dz\ell$ candidate can be 
reconstructed, the best candidate is taken to be the
one with the largest probability of converging at a single vertex. 

The tag reconstruction efficiency, extracted from signal MC and 
averaged over all data taking periods, is $(6.64\pm0.03)\times 10^{-3}$, where the 
error is due to the statistics of the signal MC sample. At this
level of selection, we find that the MC models the data
well in the electron channel of the semileptonic $B$ decay, 
but less so in the muon channel. The disagreement in the muon channel
appears to derive largely from the continuum background and 
therefore should not affect the real semileptonic tags. The
tag reconstruction efficiency is corrected for any data/MC
disagreement using a control sample described in Section \ref{sec:EextraValidation}.

\subsection{Selection of \boldmath{\btn} \xspace signal candidates}
\label{sec:SigSelection}

After the tag $B$ reconstruction, the recoil is studied for
consistency with the signal modes.
All selection criteria are optimized for each of the different signal 
$\tau$ decay modes.
The optimization is performed by maximizing the signal significance,
$s/\sqrt{s+b}$, for each channel using the signal ($s$) and background ($b$) MC and 
assuming a total branching fraction for \btn\ of $1.0 \times 10^{-4}$, using the PRIM
algorithm \cite{prim1999}. This algorithm simultaneously optimizes
selection criteria over a number of variables by relaxing and
tightening the constraints on all variables until a maximal significance
is achieved, allowing only up to a fixed percentage of signal and background
to be removed or restored with each iteration of the selection criteria.

All signal modes contain one charged particle that is identified
as either an electron, muon, or pion using standard particle identification
techniques. Both the $\tautopinu$ and the $\tautopipiznu$ modes
contain a pion signal track and are characterized by the 
number of signal $\piz$ mesons. 
The signal track
is required to have at least 12 hits in the DCH; its momentum transverse to the 
beam axis, $p_{\rm{T}}$, is required to be greater than 0.1$\gevc$, and
its point of closest approach to the interaction point must be
less than 10.0\cm along the beam axis and less than 1.5\cm transverse 
to it.
We demand the invariant mass of the signal $\piz$ be between
0.115 and 0.150$\gevcc$. The daughter photon candidates must 
have a minimum energy of 50$\mev$, and their shower shapes are required to be
consistent with electromagnetic showers.

Background consists primarily of $B^{+}B^{-}$ events in which the tag
$B$ meson has been correctly reconstructed and the recoil contains
one track and additional particles which are not 
reconstructed by the tracking detectors or calorimeter. 
These events typically contain one or more $\KL$ mesons, neutrinos and particles
that pass outside of the detector acceptance.
$\BzBzb$ and continuum events contribute background to 
hadronic $\tau$ decay modes. In addition, some excess events in data,
most likely from higher-order QED processes (such as two-photon fusion) 
that are not modeled in our MC simulation, are observed.

\begin{figure}[htb]
\begin{center}
\includegraphics[width=0.49\linewidth,keepaspectratio]{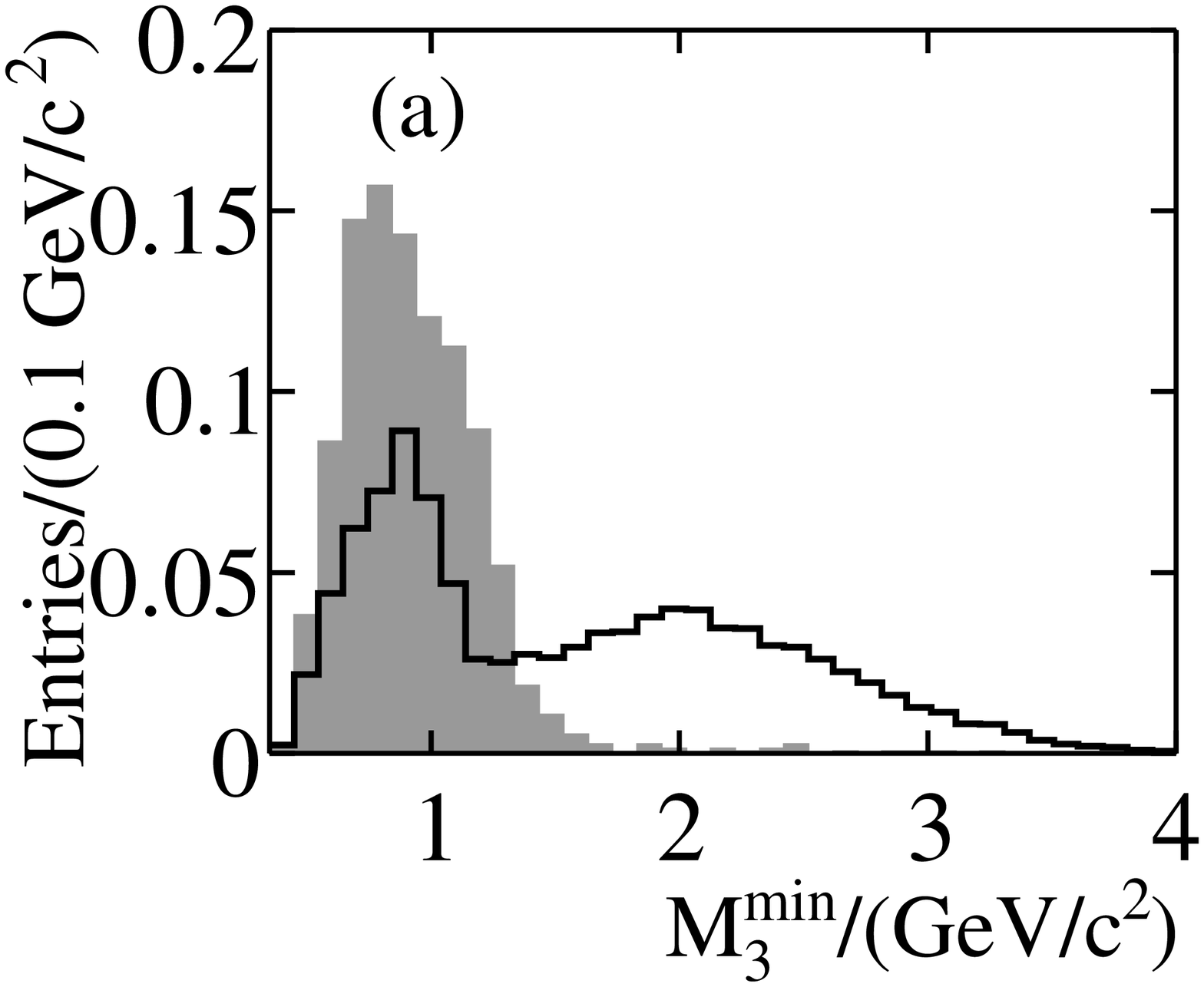}
\includegraphics[width=0.49\linewidth,keepaspectratio]{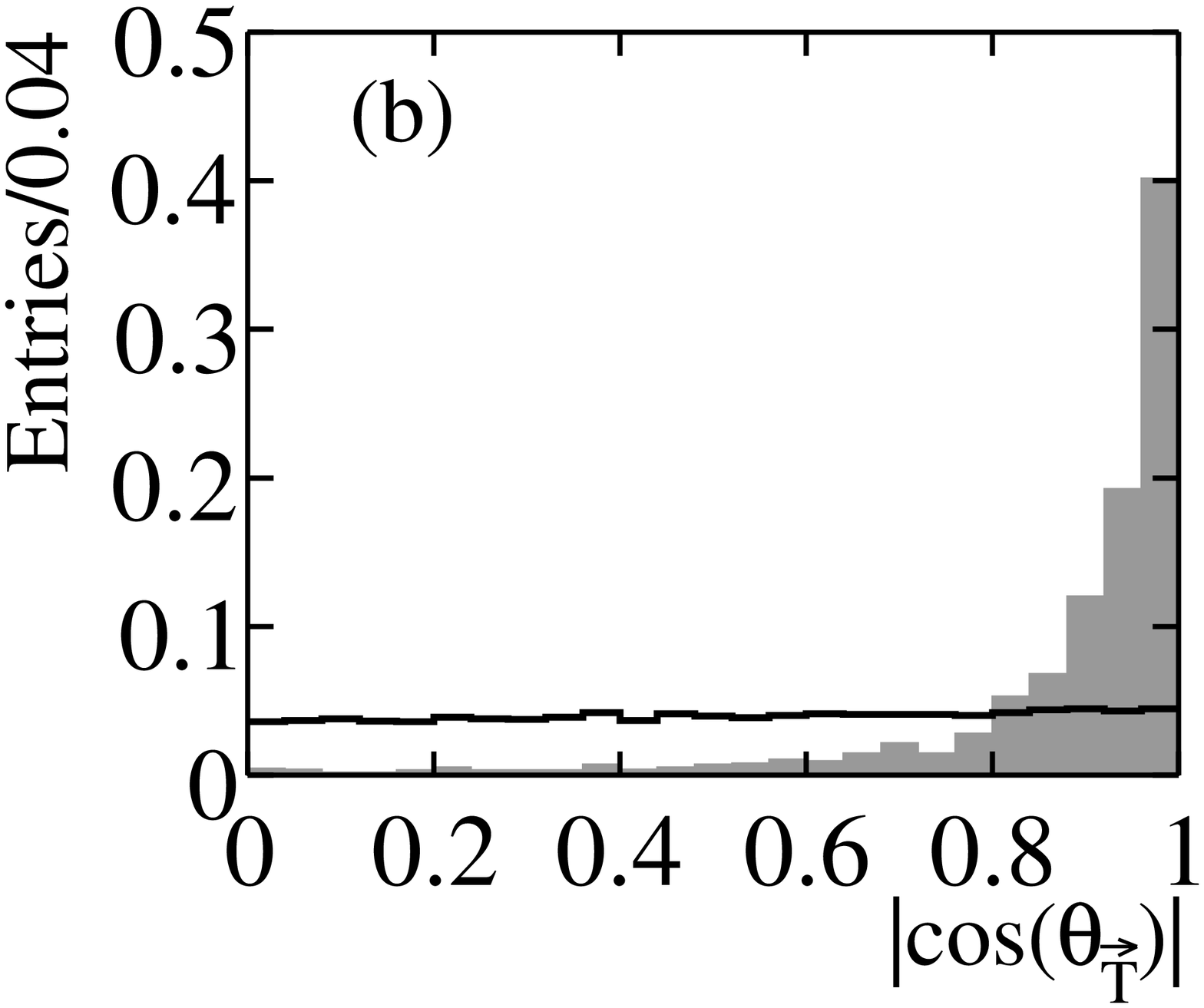}
\end{center}
\vspace{-0.8cm}
\caption{(a) Minimum invariant mass of any three tracks and (b)  
$|\cos \theta_{\vec{T}}|$ for $\btn$ signal MC (solid histogram) and $\ep\en\to\tau^+\tau^-$ MC
(gray histogram). All distributions are normalized to unit area.
}
\label{fig:tautauplot}
\end{figure}

Backgrounds are suppressed relative to signal by 
imposing constraints on the kinematic and shape properties of the events. 
The missing mass is calculated as:
\begin{equation}
M_{\rm{miss}} = \sqrt{ (E_{\FourS}-E_{\rm{vis}})^2 - ( \vec{p}_{\FourS} - \vec{p}_{\rm{vis}} )^2 }.
\end{equation}
Here ($E_{\FourS}$, $\vec{p}_{\FourS}$) is the four-momentum of the $\FourS$,
known from the beam energies. The quantities $E_{\rm{vis}}$ and $\vec{p}_{\rm{vis}}$ are the 
total visible energy and momentum of the event, which are calculated by adding the 
energy and momenta, respectively, of all the reconstructed 
tracks and photons in the event.
Continuum background is suppressed with two variables:
the cosine of the angle between the signal candidate and the tag candidate thrust
vectors (in the CM frame), $\cos \theta_{\vec{T}}$, and the minimum invariant mass 
constructible from any three tracks in an event, $M^{\rm{min}}_{3}$. 
For the background, the cosine of the thrust
angle peaks at $\pm$1, while the minimum invariant mass peaks strongly below $1.5\gevcc$, as can be seen in 
Fig.~\ref{fig:tautauplot},
where the signal and $\tau^+\tau^-$ background MC are shown. 
We project this 2-d plane into a single variable for use in the
selection optimization algorithm.
The projection uses the following empirically derived equation:
\begin{equation}
R_{{\rm cont}} \equiv \sqrt{(3.7-|\cos \theta_{\vec{T}}|)^{2} + (M^{\rm{min}}_{3}/(\rm{GeV}/c^{2})-0.75)^{2}}.
\label{eqn:rtt}
\end{equation}
Applying selection criteria to $R_{{\rm cont}}$ primarily removes background from $\ep\en\to\taup\taum$,
but also suppresses other continuum backgrounds.
Since the $\tautopipiznu$ decay proceeds via an intermediate 
resonance ($\rho^{+}\to\pip\piz$), further background rejection can be 
achieved by applying requirements on the intermediate meson candidate.
In events with more than one recoil $\piz$, the candidate with 
invariant mass closest to the nominal $\piz$ mass \cite{pdg2004}
is chosen. 
The invariant mass of the 
reconstructed $\pip\piz$ signal candidates are required to lie between
0.64 and 0.86$\gevcc$. 
A quantity analogous to $\cos \theta_{B-\Dz\ell}$,
as defined in section \ref{sec:TagReco}, can be
calculated for $\tautorho$ by replacing the $B$ with a $\tau$ and the
$\Dz\ell$ with $\pip\piz$ in Eq.~\ref{eqn:cosby}.
The analogous quantities of $|\vec{p}_{\tau}|$ and $E_{\tau}$ are calculated
assuming the $\tau$ is from the $\btn$ decay and 
that the $B^{+}$ is almost at rest in the CM frame.
We require $\cos \theta_{\tau-\pip\piz} <0.87$.

We demand that there are no $\KL$ candidates reconstructed in the IFR.
For the $\tautopi$ channel, we demand that there are fewer than two candidate clusters in
the EMC consistent with being deposited by a $\KL$.
In the leptonic final states we demand that there are two or fewer $\piz$ candidates.
For the $\tautoe$ mode, we reject events where a photon
conversion creates the electron by requiring that the invariant mass of the
signal and tag $B$ lepton pair be greater than $0.1\gevcc$.
We impose mode-dependent selection criteria on
the total momentum ($p^*_{\rm{signal}}$) of the visible decay products of the
$\tau$ candidate.

We further separate signal
and background by exploiting the remaining energy ($\eextra$),
calculated by summing the CM energy of the neutral clusters 
(with a minimum of $20\mev$ in the laboratory frame) and
tracks that are not associated with either the tag $B$ or the signal. 
Signal events tend to peak at
low $\eextra$ values whereas background events, which tend to contain
additional sources of neutral clusters, are distributed
toward higher $\eextra$ values. 
The selection applied to $\eextra$ is optimized for the best signal significance, assuming the 
branching fraction is $1\times 10^{-4}$ and was blinded for 
$\eextra < 0.5$\gev in on-resonance data until the selection 
was finalized.

The signal selection efficiencies 
for the $\tau$ decay modes are 
determined from signal MC simulation and summarized 
in Table~\ref{tab:SigSelSummary}.
The signal efficiencies correspond to the fraction of events selected 
in a specific signal decay mode, given that a tag $B$ has been reconstructed.
Signal selection efficiencies are further corrected by applying the factors
provided in Table~\ref{tab:SignalEffSys} which are explained in later sections. 

\begin{table}[!htb]
\caption{Selection criteria optimized for each signal $\tau$ decay mode.
Additional selection criteria are described in the text.
The signal efficiencies, multiplied by branching fraction, 
are given for each $\tau$ decay mode, relative to the number of tags. 
Values given in the squared brackets represent lower and upper 
selection criteria imposed on the respective quantity.
}
\centering
\begin{tabular}{lcccc} \hline
\hline
mode & $e^+$ & $\mu^+$ & $\pi^+$ & $\pi^+ \pi^{0}$ \\ \hline
$\mmiss (\rm{GeV}/c^2)$ & [4.6, 6.7] & [3.2, 6.1] & $\ge 1.6$  &   $\le 4.6$  \\ 
$p^{*}_{\rm{signal}} (\rm{GeV}/$c$)$ & $\le 1.5$ &   --  & $\ge 1.6$  & $\ge 1.7$      \\ 
$R_{{\rm cont}}$ &  [2.78, 4.0] & $> 2.74$ & $> 2.84$ & $> 2.94$  \\
$\eextra (\rm{GeV})$   & $< 0.31$ & $< 0.26$ & $< 0.48$ &  $< 0.25$ \\ 
\hline
Efficiency (\%)    &  $4.2\pm 0.1$ & $2.4\pm 0.1$ & $4.9\pm 0.1$ & $1.2\pm 0.1$ \\
\hline \hline
\end{tabular}
  \label{tab:SigSelSummary}
\end{table}

\subsection{Background Estimation from \boldmath{\eextra} Sidebands}
\label{sec:EextraSBExtrapolation}

We estimate our background from the data by studying events
in a sideband region of $\eextra$. We define the sideband (sb) region
as $\eextra>0.5\gev$, and also define signal regions (sig) in $\eextra$
using the appropriate signal mode-dependent selection. After applying all other
selection criteria, we compute from the background MC simulation the ratio of events in the
sideband ($N_{\mbox{\scriptsize{MC,sb}}}$) 
and signal ($N_{\mbox{\scriptsize{MC,sig}}}$) regions,
\begin{eqnarray}
R^{\mbox{\scriptsize{MC}}} & = & \frac{N_{\mbox{\scriptsize{MC,sig}}}}{N_{\mbox{\scriptsize{MC,sb}}}}.
\end{eqnarray}
\noindent Using the number of data events in the sideband ($N_{\mbox{\scriptsize{data,sb}}}$)
and the ratio $R^{\mbox{\scriptsize{MC}}}$, the number of expected background events in the 
signal region in data ($N_{\mbox{\scriptsize{exp,sig}}}$) is estimated,

\vspace{-0.5cm}
\begin{eqnarray}
N_{\mbox{\scriptsize{exp,sig}}} & = & N_{\mbox{\scriptsize{data,sb}}} \cdot R^{\mbox{\scriptsize{MC}}}.
\end{eqnarray}

\noindent
The sideband background projection (Table \ref{tab:validation})
is taken as the number of expected background events.

\begin{table}[!htb]
\caption{Comparison of the expected total background, computed from data and MC
in the $\Dz$ mass sideband and signal regions, to that computed by projecting the 
$\eextra$ sideband into the $\eextra$ signal region.}
\centering
\begin{tabular}{ccccc} 
\hline
\hline
       & \multicolumn{4}{c}{Background Prediction}\\
signal mode & $e^+$ & $\mu^+$ & $\pi^+$ & $\pi^+ \pi^{0}$ \\
\hline\\[-6pt]
$\eextra$ sideband  & 44.3$\pm$5.2 & 39.8$\pm$4.4 & 120.3$\pm$10.2 & 17.3$\pm$3.3 \\
$\Dz$ sideband      & 44.2$\pm$6.4 & 42.8$\pm$6.0 & 113.4$\pm$11.6 & 16.3$\pm$4.5  \\
\hline \hline
\end{tabular}
  \label{tab:validation}
\end{table}

The background estimate is validated by performing a similar test using sidebands
in the $\Dz$ mass distribution. We select events using \Dz\ mass sidebands
between $4\sigma$ and $9\sigma$ above and below the nominal \Dz\ mass, with all other
signal selection criteria applied. Candidates in these regions of the \Dz\ mass
distribution are random combinations of kaons and pions, and represent
a pure combinatoric background. We average the yields from the upper and lower
sidebands and scale this using the ratio of the \Dz\ mass sideband and signal region.
This yields a \Dz\ mass combinatoric background 
estimate in the \Dz\ mass signal region 
for both data ($N_{{\rm comb}}^{{\rm data}}$) and MC ($N_{{\rm comb}}^{{\rm MC}}$). 
The remaining component, in the MC, of the background which contains real \Dz\ mesons
in the tag is then computed,
\begin{equation}
N_{{\rm peak}}^{{\rm MC}} = N_{{\rm total}}^{{\rm MC}} - N_{{\rm comb}}^{{\rm MC}},
\end{equation}
and added to the combinatoric component (determined from data) to obtain an 
effective estimate of the total background,
\begin{equation}
N_{{\rm total}}^{{\rm predicted}} = N_{{\rm peak}}^{{\rm MC}} + N_{{\rm comb}}^{{\rm data}}.
\end{equation}
This is done for each reconstructed signal decay channel.
The method assumes that the background in the $\eextra$ signal region can be 
modeled by the combinatoric component of the \Dz\ mass distribution, taken from
data, and the peaking component of the \Dz\ mass distribution, taken from 
MC simulations. Since it uses the \Dz\ mass sidebands, it is also statistically
independent from the $\eextra$ sideband calculation.

We find very good agreement between the background
prediction using the \Dz\ mass sidebands and that obtained from the projection 
of the $\eextra$ sideband. This agreement is demonstrated in
Table \ref{tab:validation} and further validates our background
estimation method.

\subsection{Correction of tag \boldmath{$B$} yield and \boldmath{\eextra} 
simulation}
\label{sec:EextraValidation}

The tag $B$ yield and $E_{\rm{extra}}$ distribution in signal and background MC simulation 
are validated using control samples. These samples are further used to define corrections to
efficiencies of selection criteria.
``Double-tagged'' events, for which
both of the $B$ mesons are reconstructed in tagging modes, 
$\btodlnux$~vs. $\B^+ \to \Dzb \ell^+ \nul X$ are used as the primary control sample. 
``Single-tagged'' events are also used where one $B$ decays via 
$\btodlnux$ and the other $B$ decay is not constrained. 
The double-tagged sample is almost entirely free of continuum events.

\begin{figure}[htb]
\begin{center}
\includegraphics[width=0.8\linewidth,keepaspectratio]{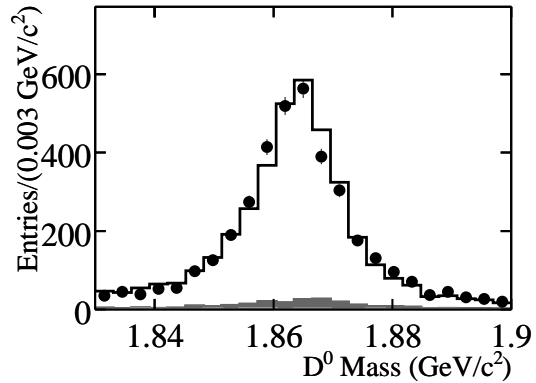}
\end{center}
\vspace{-0.8cm}
\caption{$\Dz$ invariant mass from the recoil $B$ meson in double-tagged
events.  
On-resonance data (black circles) are overlaid on the 
combined \BB\ (solid histogram) and continuum (gray histogram) MC samples normalized to the data luminosity.}
\label{fig:d0Mass_DoubleTag}
\end{figure}

We select double-tagged events by requiring that
the two semileptonic $B$ candidates have opposite charge and do not share 
any particles. We also require that there are no additional tracks in the event. 
If there are more than two such 
independent tag $B$ candidates in the event then the two
best candidates are selected as those with the largest probabilities of
each converging at a common origin. The $\Dz$ meson invariant mass is shown 
in Fig.~\ref{fig:d0Mass_DoubleTag} for the second tag in all double-tagged events.

We initially determine the tag efficiency using 
a signal MC where one of the two $\B$ mesons always decays into a generic
final state and the other always decays into a $\taup\nu$ final state. 
We estimate the correction to the MC semileptonic tag efficiency by comparing the
number of single- and double-tagged events in data and MC. We calculate the ratio of
double-tagged to single-tagged events, and we use the ratio of this quantity from data and MC 
as a correction factor for the tag $B$ yield.

We determine the number of single-tagged events by subtracting the combinatoric component under 
the $\Dz$ mass peak in events where one \B\ is tagged and the second is allowed to
decay without constraint (Fig. \ref{fig:d0masstag_pstartaglep}). 
We determine this component by using \Dz\ mass sidebands between $4\sigma$
and $7\sigma$ above and below the nominal \Dz\ mass. A narrower sideband region is used
for this study than in the background estimate validation due to the comparative flatness of the sidebands in this region
and the large statistics available at this early stage of selection. We then average the yields from these
combinatoric \Dz\ mass regions and scale by the ratio of the sideband and signal region
widths. We perform this subtraction using events where the \Dz\ meson from one of the semileptonic tags is
reconstructed as $\Dz \to K^{-} \pip$ and the second tag decays into any of our allowed
final states. The resulting single-tagged event yields, and the double-tagged event yields,
are shown in Table~\ref{tab:double_single_tag_yields}. We compared these results
to that obtained from events where the \Dz\ of at least one of the tags decays as $\Dz \to K^{-} \pip \pim \pip$
and found a similar correction.

We take the uncertainty on the data/MC single-to-double-tag ratios
as the systematic uncertainty on the tag $B$ yield. We find a correction of
$1.05$ with a $3.6\%$ uncertainty.
This comparison between data and MC provides a more realistic environment than  
signal MC in which to compare the various forms of background in the analysis, 
and correct for them. The double-tagged sample alone would only correct for 
$B^+B^-$ backgrounds.

\begin{table}[!htb]
\caption{\label{tab:double_single_tag_yields}Single-tag and double-tag yields in data and MC, for events
where the \Dz\ meson from the first tag is required to decay as $\Dz \to K^{-}\pip$. 
We calculate the ratio of these two yields,
and take the ratio of these ratios as a correction to the tagging efficiency determined from
MC. The uncertainty on the correction is taken as a systematic error.}
\begin{center}
\begin{tabular}{lccc}
\hline\hline\\[-6pt]
\hspace{0.2cm}&
Single Tags&
Double Tags&\vspace{0.1cm}
Ratio \tabularnewline
\hline
Data \hspace{0.2cm}&
$335417\pm747$&
$1067 \pm 33$&
($3.18\pm0.10$)$\times 10^{-3}$\tabularnewline
MC \hspace{0.2cm}&
$349972 \pm 572$&
$1065 \pm 20$&
($3.04\pm0.06$)$\times 10^{-3}$ \tabularnewline
\multicolumn{2}{l}{Data/MC} &  & $1.049\pm0.038$\tabularnewline
\hline\hline
\end{tabular}\par\end{center}
\end{table}

\begin{figure}[htb]
\begin{center}
\includegraphics[width=0.8\linewidth,keepaspectratio]
{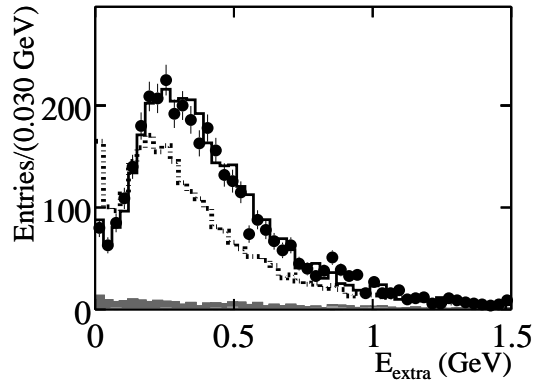}
\end{center}
\vspace{-0.8cm}
\caption{$\eextra$ after the reconstruction of two non-overlapping semileptonic $B$ candidates.
On-resonance data (black circles) are overlaid on the 
combined \BB\ (solid histogram) and continuum (gray histogram) MC samples normalized to the data luminosity. $\btn$ signal MC (dashed-dotted line) is shown for comparison, with arbitrary normalization.
}
\label{fig:eextra_DoubleTag}
\end{figure}

We can further test the modeling of $\eextra$ by comparing
it in double-tagged events from data and MC. 
The $E_{\rm{extra}}$ for the double-tagged sample (Fig.~\ref{fig:eextra_DoubleTag})
is calculated by summing the 
energy of the photons which are not associated with 
either of the tag $B$ candidates. The sources of photons contributing
to the $E_{\rm{extra}}$ distribution in double-tagged events  
are similar to those contributing to the $E_{\rm{extra}}$ distribution  
in the signal MC simulation.

We additionally check the modeling of $E_{\rm{extra}}$ by comparing samples of 
events where the signal and
tag $B$ candidates are of the same sign. 
We find that for all signal modes, there is good agreement between the
shape of the $E_{\rm{extra}}$ from the background prediction and the data
 in this wrong-charge sample. In the pion channel in particular, we find the
data yield is higher than predicted from MC. This suggests that for a pure background sample, 
with a topology similar to that of signal, the $\eextra$ distribution is well-modeled
but the background estimate cannot be taken directly from the MC background simulation. 
This further validates our choice to take the background estimate from 
the $E_{\rm{extra}}$ sideband in data and the signal-to-sideband ratio in MC
simulation.

\section{Studies of Systematic Uncertainties}
\label{sec:Systematics}

The main sources of uncertainty in the determination of the $\btn$
branching fraction are the tag reconstruction efficiency
($\varepsilon_{\rm{tag}}$),
the efficiency of each signal mode ($\varepsilon_{i}$),
and the number of expected background
events in the signal region for each signal mode.

An uncertainty of 1.1\% enters the branching fraction calculation 
from the estimation of the number of $B^{+}B^{-}$ events present in the 
data sample \citep{BCount}.
The tagging efficiency and yield in signal MC is corrected 
using the double-tagged and single-tagged samples. 
The tag $B$ yield systematic uncertainty is 3.6\%, with a correction factor
to the yield of 1.05.
The systematic uncertainties on the signal efficiency depend on the $\tau$ decay mode 
and include effects such as the tracking of charged particles,
particle identification, and the modeling of $\piz$ mesons. 

The systematic uncertainty on the signal efficiency due to the mis-modeling 
of the $\eextra$ variable is extracted using the double-tagged events. 
We extract the yield of candidates satisfying $\eextra <$~0.5\gev. 
This yield is then compared to the number of candidates in the full sample.
Comparing the ratio extracted from MC to that extracted from data yields a correction factor, 
the error of which is taken as the systematic uncertainty for $\eextra$. 
The systematic uncertainty for $\eextra$ is 3.4\% with a correction of 0.99.

\begin{table}
\caption{Contributions to the systematic uncertainty (in percent) on the signal selection efficiencies 
for different selection modes. The total summed uncertainty is added linearly with the
systematic uncertainties from IFR $\KL$ reconstruction and $\eextra$ modeling.
The result of this (``signal $B$'') is added together in quadrature with the 
uncertainty on tag $B$ reconstruction and the number of \BB\ pairs 
in the sample ($N_{\BB}$). The ``Correction Factor'' is a multiplicative factor applied to the 
efficiency for each mode.}
\begin{center}
\begin{tabular}{lcccc} \hline \hline
$\tau$ decay mode    & $\enunu$   & $\mununu$  & $\pinu$  & $\pipiznu$ \\
\hline
Tracking                  & 0.5 & 0.5 & 0.5 & 0.5 \\
Particle Identification   & 2.5 & 3.1 & 0.8 & 1.5 \\
$\piz$                    & --  & --  & --  & 2.9 \\
EMC $\KL$                 & --  & --  & 3.8 & --  \\
\hline
IFR $\KL$                 & \multicolumn{4}{c}{3.3} \\
$\eextra$                 & \multicolumn{4}{c}{3.4} \\
\hline
signal $B$                & \multicolumn{4}{c}{5.5} \\
tag $B$                   & \multicolumn{4}{c}{3.6} \\
$N_{\BB}$                 & \multicolumn{4}{c}{1.1} \\

\hline
Total                     & \multicolumn{4}{c}{6.6} \\
\hline
\hline
Correction Factor         & 0.951 & 0.868 & 0.964 & 0.939 \\
\hline
\end{tabular}
\end{center}
\label{tab:SignalEffSys}
\end{table}

The systematic uncertainty on the modeling of $\KL$ candidates is extracted  
using the double-tagged events, similar to the method used for the $\eextra$ 
systematic evaluation. 
We quantify the data/MC comparison by extracting the yield with a cut demanding 
exactly zero (less than two) reconstructed 
IFR (EMC) measured $\KL$ candidates remaining, and extracting the yield with a sample where any number of $\KL$ candidates 
remain, and take the ratio of ratios from the MC and data.
The systematic uncertainty for vetoing IFR (EMC) $\KL$ candidates is 3.3\% (3.8\%), with a 
correction factor on the efficiency of 0.99 (0.97).

A breakdown of the contributions
to the systematic uncertainty for each signal mode is given in Table~\ref{tab:SignalEffSys}.
We find that the most significant individual effects on the signal efficiency
are from the modeling of the $\eextra$ and the $\KL$ vetos. The uncertainties on each mode
are combined by weighting them by the corrected efficiency for a given mode, using the
efficiencies from Table~\ref{tab:SigSelSummary} multiplied by the correction factors given in 
Table~\ref{tab:SignalEffSys}. The signal-mode-specific systematic uncertainties are summed
in quadrature and then the sum is added linearly with the IFR \KL\ and $\eextra$ 
uncertainties, which are correlated among the modes. The resulting overall systematic uncertainty on
the signal efficiency is then added in quadrature with the uncertainties on the 
tag $B$ reconstruction and the number of \BB\ pairs in the sample
to give a total uncertainty of 6.6\%.

\section{Results}
\label{sec:Physics}

\begin{figure}[htb]
\begin{center}
\includegraphics[width=.35\textwidth]{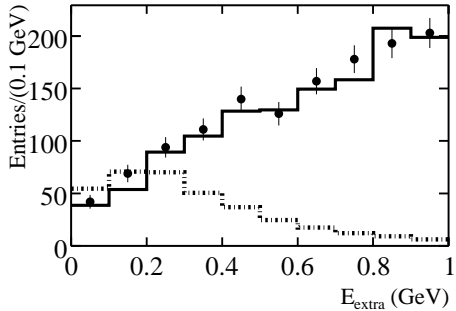}
\end{center}
\vspace{-0.8cm}
\caption{$\eextra$ distribution. All selection criteria 
have been applied and all signal modes combined. 
Background MC (solid histogram) has been normalized to the luminosity 
of the on-resonance data (black dots), and then
additionally scaled according
to the ratio of predicted background from data and MC as 
presented in section~\ref{sec:EextraSBExtrapolation}.
$\btn$ signal MC (dotted histogram) is normalized to a branching
fraction of $10^{-3}$ and shown for comparison.}
\label{fig:eextra_allcuts_ALL}
\end{figure}

\begin{figure}[htb]
\begin{center}
\includegraphics[width=1.0\linewidth]{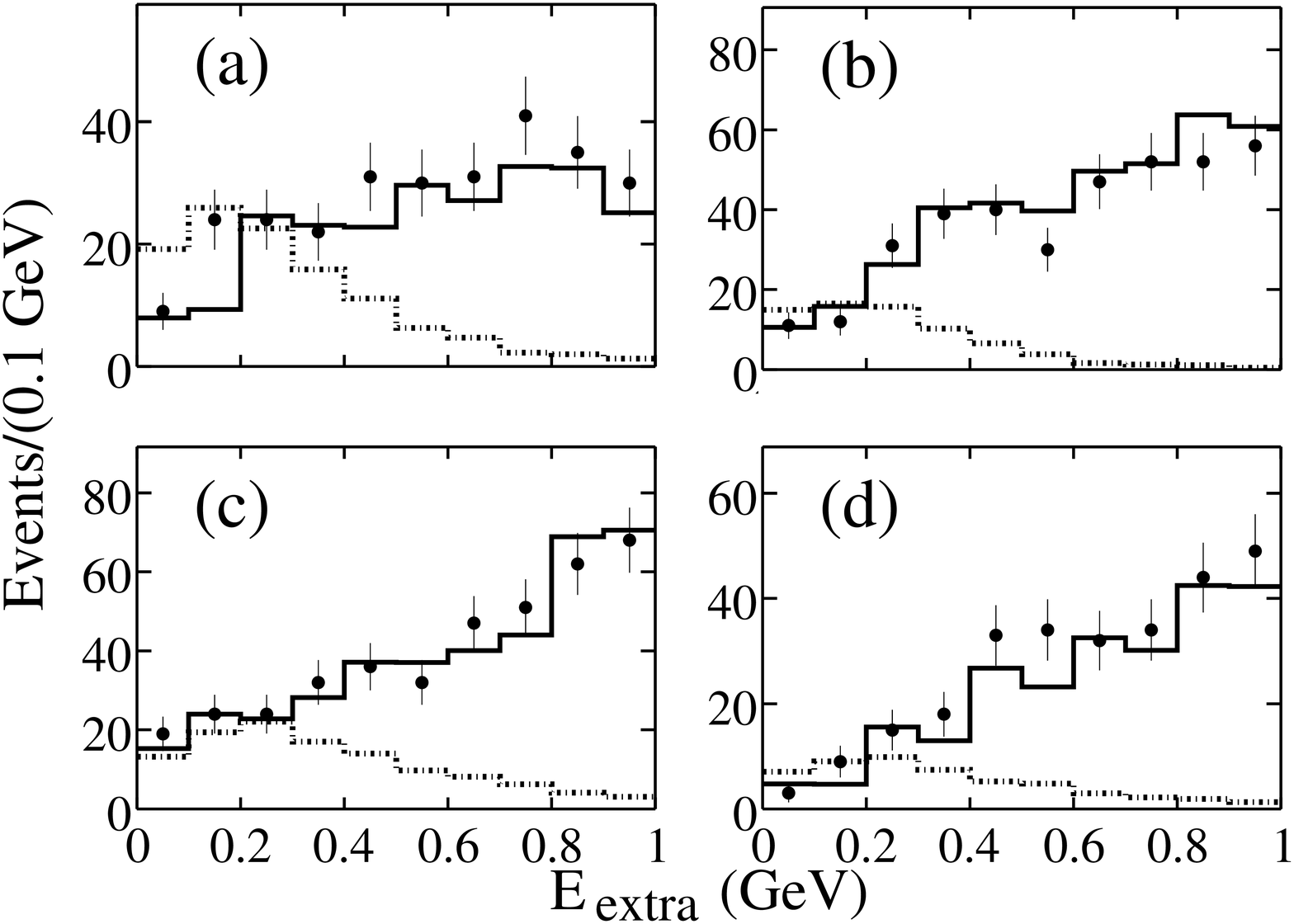}
\end{center}
\vspace{-0.8cm}
\caption{$\eextra$ distribution after all selection criteria 
for (a) $\tautoenunu$, (b) $\tautomununu$, (c) $\tautopinu$, and 
(d) $\tautopipiznu$.
Background MC (solid histogram) has been normalized to the luminosity 
of the on-resonance data (black dots), and then
additionally scaled according
to the ratio of predicted background from data and MC as 
presented in section~\ref{sec:EextraSBExtrapolation}.
$\btn$ signal MC (dotted histogram) is normalized to a branching
fraction of $10^{-3}$ and shown for comparison.}
\label{fig:eextra_allcuts_Separated}
\end{figure}

After finalizing the signal selection criteria, we measure the
yield of events in each channel in the signal region
of the on-resonance data. 
Table~\ref{tab:unblind-result} lists the
number of observed events in on-resonance data in the signal region,
together with the expected number of background events in the 
signal region (taken from the $\eextra$ sideband prediction from Table~\ref{tab:validation}). 
Figure~\ref{fig:eextra_allcuts_ALL} shows the $\eextra$
distribution for all data and MC in the signal region, with signal MC shown for
comparison. Figure~\ref{fig:eextra_allcuts_Separated} shows the $\eextra$ distribution
separately for each of the signal modes.
\begin{table}[hbt]
\centering
\caption{\label{tab:unblind-result}
Observed number of on-resonance data events in the signal region are shown, 
together with number of expected background events.}
\begin{tabular}{lcc} \hline \hline
$\tau$            & Expected background  &  Observed events  \\ 
decay mode        & events               &  in on-resonance data  \\ \hline
$\tautoenunu$     & 44.3  $\pm$ 5.2   & 59  \\ 
$\tautomununu$    & 39.8  $\pm$ 4.4   & 43  \\ 
$\tautopinu$      & 120.3 $\pm$ 10.2  & 125  \\ 
$\tautopipiznu$   & 17.3  $\pm$ 3.3   & 18  \\ 
\hline
All modes    & 221.7 $\pm$ 12.7  & 245  \\ \hline \hline
\end{tabular}
\end{table}

We determine the \btn \xspace branching fraction from the number of signal
candidates $s_i$ in data for each $\tau$ decay mode, according to $s_i =
N_{\BB} \mathcal{B}(\btn) \varepsilon_{\rm{tag}} \varepsilon_i$, where
$N_{\BB}$ is the total number of \BB\ pairs in data.
The results from each of our four signal decay channels ($n_{\rm ch}$)
are combined using the estimator
$Q = {\cal L}(s+b)/{\cal L}(b)$,
where ${\cal L}(s+b)$ and ${\cal L}(b)$ are the
likelihood functions for signal plus background and background-only
hypotheses, respectively:
\begin{equation}
  {\cal L}(s+b) \equiv
  \prod_{i=1}^{n_{\rm ch}}\frac{e^{-(s_i+b_i)}(s_i+b_i)^{n_i}}{n_i!},
        \;
  {\cal L}(b)   \equiv
  \prod_{i=1}^{n_{\rm ch}}\frac{e^{-b_i}b_i^{n_i}}{n_i!}.
  \label{eq:lb}
\end{equation}

\begin{figure}
\begin{center}

\vspace{0.3cm}
\includegraphics[width=0.8\linewidth,keepaspectratio]{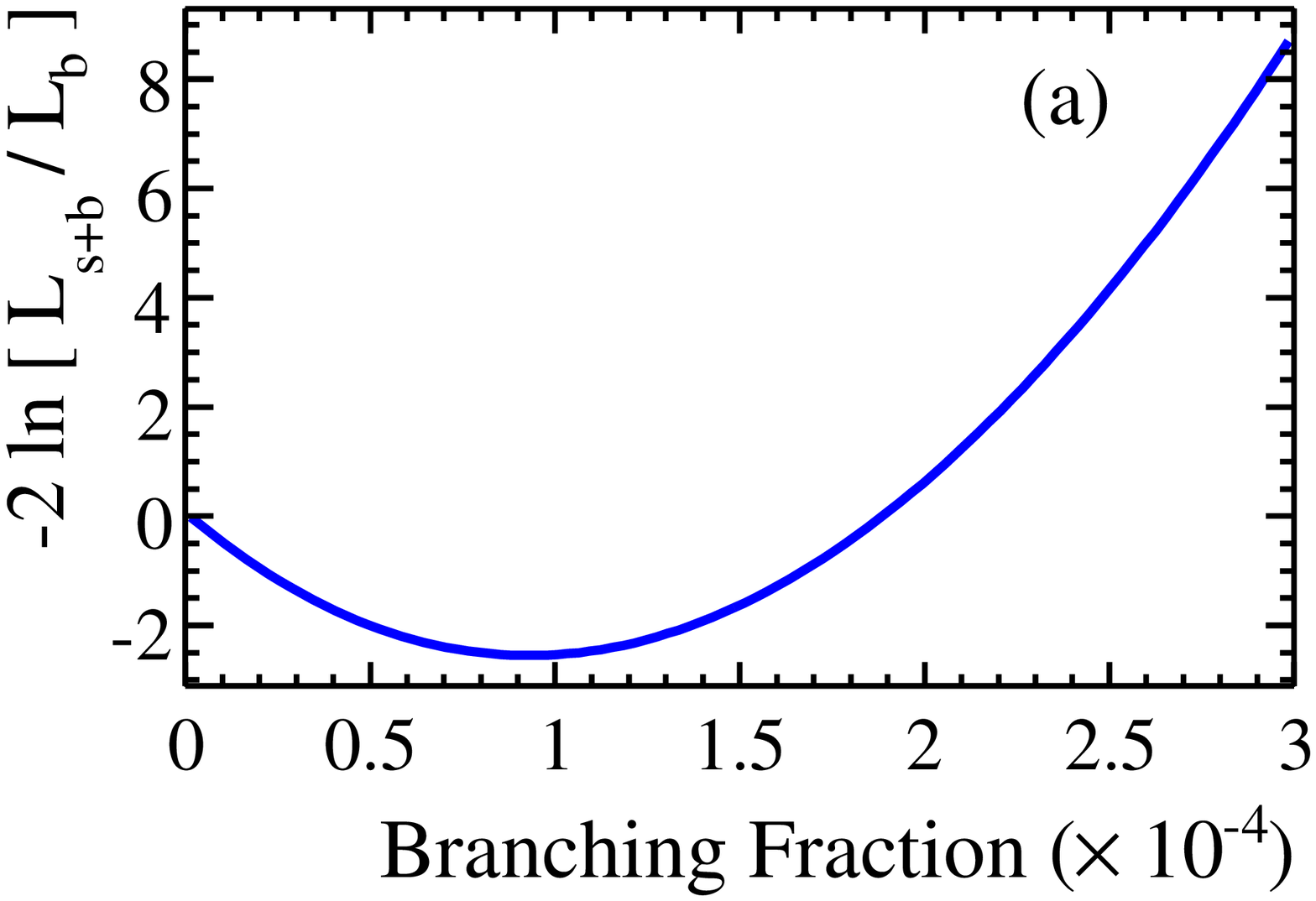}

\includegraphics[width=0.8\linewidth,keepaspectratio]{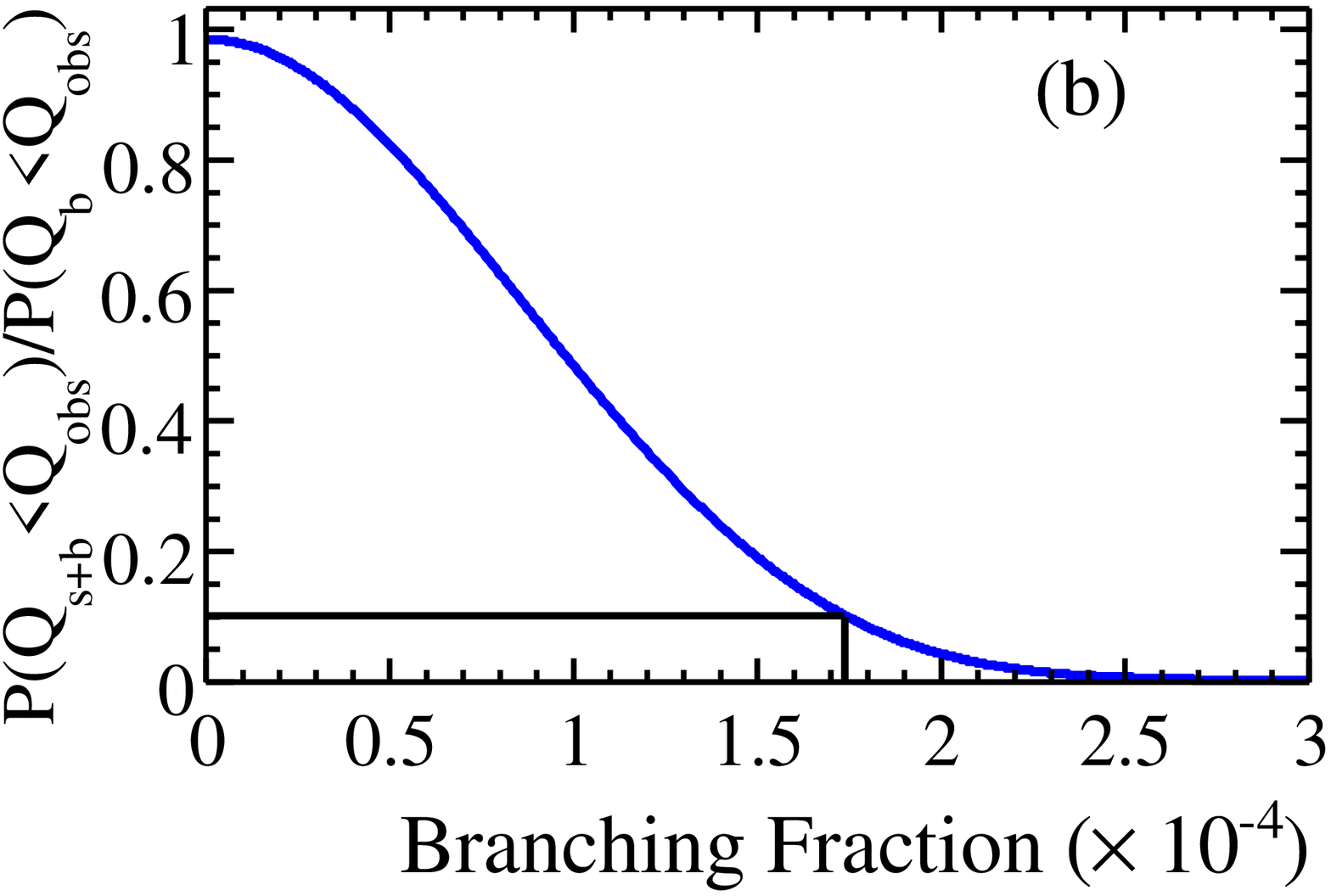}
\end{center}
\vspace{-0.30cm}
\caption{(a) Twice the negative natural logarithm of the likelihood ratio as a function of
signal branching fraction hypothesis and (b) the upper limit as a function of
signal branching fraction hypothesis (where the
horizontal and vertical intersecting lines indicate the 90\% CL limit).
}
\label{fig:unblindCL}
\end{figure}

\noindent We include the systematic uncertainties, including those of a
statistical nature, on the expected
background ($b_{i}$) in the likelihood definition by
convolving it with a Gaussian function.
The mean of the Gaussian is $b_{i}$, and
the standard deviation ($\sigma_{b_{i}}$) of the Gaussian is the  
error on $b_i$~\cite{lista}.

We calculate the branching fraction central value (including statistical uncertainty and uncertainty
from the background) by scanning
over signal branching fraction hypotheses between $0.0$ and $3.0 \times 10^{-4}$ in steps
of $0.025\times10^{-4}$ and computing the value of $\mathcal{L}(s+b)/\mathcal{L}(b)$ for each
hypothesis (Fig. \ref{fig:unblindCL}a). 
The branching fraction is the hypothesis which minimizes $-2 \log(\mathcal{L}(s+b)/\mathcal{L}(b))$,
and the statistical uncertainty is determined by finding the points on the likelihood scan that
occur at one unit above the minimum. The systematic error is determined
as detailed in section \ref{sec:Systematics} and computed for the branching fraction as 
a fraction of the central value.

The upper limit at the 90\% CL, including both statistical and systematic uncertainties,
is determined by generating 5000 experiments
for each of the aforementioned signal branching fraction hypotheses. 
Each generated experiment also includes the expected number of background
events, and varies the generated number of background in each channel according to
its uncertainty. The total number 
of events is allowed to vary according to Poisson statistics, and systematics are incorporated
in the efficiency for each channel and the number of $B$ mesons originally produced
by the collider. The number of signal events in each channel (labeled $i$)
for each experiment is thus computed from the branching fraction hypothesis as:
\begin{equation}
s_{i} = \mathcal{B}_{i} \times \mathcal{G}(\varepsilon_{i},\delta\varepsilon_{i}) \times \mathcal{G}(N_{\Bp},\delta N_{\Bp}),
\end{equation}
where $\mathcal{G}(x,\delta x)$ represents a number sampled from a Gaussian distribution 
centered on the quantity
$x$ with systematic uncertainty $\delta x$, and $\varepsilon_{i}$ and $N_{\Bp}$ are the efficiency in each
channel and the number of charged $B$ mesons produced by the collider, respectively. 
Each experiment therefore contains a generated number of signal and a generated
number of background which will vary around the input hypotheses $s$ and $b$ according
to the above procedures.

We determine the confidence level of a given signal hypothesis by finding the probability
that the value of the estimator $Q$ in experiments generated according to a given composition
(signal and background, $Q_{s+b}$,  or only background, $Q_{b}$) 
is less than that observed in data ($Q_{\rm obs}$).
The 90\% CL limit is determined by using the CLs method~\cite{cls}, in which we determine 
the signal hypothesis for
which $P(Q_{{\rm s+b}} < Q_{{\rm obs}})/P(Q_{{\rm b}} < Q_{{\rm obs}}) = 1 - 0.90$, 
where $P(Q_{{\rm s+b}} < Q_{{\rm obs}})$ ($P(Q_{{\rm b}} < Q_{{\rm obs}})$) is the probability 
that experiments generated assuming a given $s+b$ ($b$) hypothesis
have a likelihood ratio lower than that observed in data (Fig. \ref{fig:unblindCL}b).

We determine the branching fraction central value to be
\begin{equation}
\mathcal{B}(\btn) = (0.9\pm{0.6}(\mbox{stat.}) \pm 0.1 (\mbox{syst.})) \times 10^{-4}
\label{eqn:bf}
\end{equation}
and set an upper limit at the 90\% CL of
\begin{equation}
\mathcal{B}(\btn) < 1.7 \times 10^{-4}. 
\label{eqn:ul}
\end{equation}
The central value of the branching fraction is in agreement with that measured by
the Belle Collaboration at the level of two standard deviations. 
We interpret this result in the context of the Standard Model.
Using the central value for $\mathcal{B}(\btn)$ and taking the known values of
$G_F$, $m_B$, $m_{\tau}$ and $\tau_{B}$~\cite{pdg2004} we calculate, from Eq.~\ref{eqn:br}, 
$f_{B}\cdot|\Vub| = (7.2^{+2.0}_{-2.8}(\mbox{stat.})\pm{0.2}(\mbox{syst.}))\times10^{-4}$\gev, where the uncertainties are non-Gaussian.
Using the value of $|\Vub|$ from~\cite{pdg2004} we extract 
$f_B = 0.167^{+0.048}_{-0.066}\gev$, where the uncertainty is dominated by the statistical
uncertainty on the branching fraction central value. \\

\section{Acknowledgments}
\label{sec:Acknowledgments}

\vspace{-0.2cm}
We are grateful for the 
extraordinary contributions of our \pep2\ colleagues in
achieving the excellent luminosity and machine conditions
that have made this work possible.
The success of this project also relies critically on the 
expertise and dedication of the computing organizations that 
support \babar.
The collaborating institutions wish to thank 
SLAC for its support and the kind hospitality extended to them. 
This work is supported by the
US Department of Energy
and National Science Foundation, the
Natural Sciences and Engineering Research Council (Canada),
Institute of High Energy Physics (China), the
Commissariat \`a l'Energie Atomique and
Institut National de Physique Nucl\'eaire et de Physique des Particules
(France), the
Bundesministerium f\"ur Bildung und Forschung and
Deutsche Forschungsgemeinschaft
(Germany), the
Istituto Nazionale di Fisica Nucleare (Italy),
the Foundation for Fundamental Research on Matter (The Netherlands),
the Research Council of Norway, the
Ministry of Science and Technology of the Russian Federation, 
Ministerio de Educaci\'on y Ciencia (Spain), and the
Particle Physics and Astronomy Research Council (United Kingdom). 
Individuals have received support from 
the Marie-Curie IEF program (European Union) and
the A. P. Sloan Foundation.

\bibliography{btaunu}

\begin{thebibliography}{17}
\bibitem[{cc()}]{cc}
\bibinfo{note}{Charge-conjugate modes are implied throughout the paper.}
\end{thebibliography}

\end{document}